\documentclass[10pt,a4paper,twoside]{article}

\usepackage{indentfirst}                        % Required!
\newenvironment{resum}{\begin{quote}\small}{\end{quote}}

\newcommand{\bfsf}[1]{\textsf{\textbf{#1}}}

\def\Journal#1#2#3#4#5#6{{#1} {\bf #2}, #3 (#5)}
\def\CQG{\it Class.\ Quantum\ Grav.\/}
 \def\CMP{\em Commun. Math. Phys.}

\def\MPL{\it Mod.\ Phys.\ Lett.\/}
\def\mm2{{\cal V}}%(halves with boundary)(OLD complete ones)

\def\d{{\mbox d}}

\newcommand{\bm}[1]{\mbox{\boldmath $#1$}}
\def\spI{(\mm2_I, g_I)}
\def\spE{(\mm2_E, g_E)}

\def\kk{\kappa}
\def\vkk{\vec \kappa}
\def\lie{{\cal L}}

\def\ernst{{\cal E}}
\def\jj{{\cal J}}
\def\Tr{{\mbox{Tr}}}
\def\sign{\mbox{sign}}
\def\id{\mbox{1\hspace{-2.8pt}I}}
\def\jjj{\underline{\jj}}
\def\Ima{\mbox{Im}}

\begin{document}

\thispagestyle{plain}           % Remove headings in first page

\begin{center}

%  Title; use linebreaks with "\\" if necessary:

{\LARGE\bfsf{On the construction of global models\\
describing isolated
rotating charged bodies;\\
uniqueness of the exterior
gravitational field}}

\bigskip

%  Author list:

\textbf{Ra\"ul Vera}

%  Affiliation. State only name and Institution of authors:

\textsl{Dublin City University, Ireland.} \\

\end{center}

\medskip

%  Place abstract here:

\begin{resum}
A relatively recent study by Mars and Senovilla
provided us with a uniqueness result
for the exterior vacuum gravitational field
of global models describing finite isolated rotating bodies
in equilibrium in General Relativity (GR).
The generalisation to exterior electrovacuum
gravitational fields, to include
charged rotating objects, is presented here.
\end{resum}

\bigskip

%  ********************************************************
%               Contribution proper begins here:
%  ********************************************************

\section{Introduction}
The description of astrophysical self-gravitating isolated, rotating
bodies in equilibrium in GR is still poorly understood.
To describe such bodies we would require global models for
finite isolated rotating objects in equilibrium, but we still lack complete
non-spherically symmetric models for a finite body
together with its exterior.

The global models one is interested in
consist of stationary spacetimes,
to account for the equilibrium state, being also asymptotically
flat, to account for the isolation of the body. Furthermore, it has
been always argued that for the model to be
in a final equilibrium state, the spacetime has to be axisymmetric.
Now, the framework used here lies on the construction
of such global models by means of the matching of spacetimes:
%for the description of the finite object:
one spacetime $\spI$
describing the interior of the body and another describing the
exterior $\spE$, matched across a hypersurface that
describes the surface of the body at all times.
The global model is then theoretically attacked
by solving the corresponding Einstein field equations at both
the interior and the exterior taking the common boundary data
at the matching (timelike) hypersurface $\sigma$,
which is assumed to preserve the symmetry~\cite{mps}.
Regarding the {\it vacuum} exterior problem, uniqueness
for a stationary axially symmetric asymptotically flat
vacuum solution with boundary conditions at $\sigma$ coming
from the matching with an interior region was solved in \cite{MASEuni}.
%EXISTENCE?
The aim of this work has been to generalize the uniqueness results
for the vacuum exterior to exteriors containing electromagnetic fields
without sources, i.e. {\it electrovacuum} solutions of the Einstein-Maxwell
equations,
in order to describe the exterior gravitational field of
finite (and spatially simply-connected) charged bodies.
As a natural further assumption, the exterior electromagnetic (e-m)
field
is supposed to be also stationary and axisymmetric.

In the following, $*$ denotes the Hodge dual, and
$\wedge$ the exterior product.
%, $\Tr$ the trace
%and ${}^+$ the 

\section{The electrovac exterior; the exterior problem}
Let us consider a strictly stationary (free of ergoregions and/or
Killing horizons) and axisymmetric
electrovacuum spacetime $\spE$ (containing a stationary
and axisymmetric e-m field). There
exists a coordinate system $\{t,\phi,\rho,z\}$,
the so-called Weyl coordinates, in which the corresponding
line-element reads globally
\begin{equation}
  \label{eq:ds2e}
  \d s^2_E=-e^{2U}\left(\d t+ A \d\phi\right)^2+e^{-2U}
\left[e^{2k}\left(\d\rho^2+\d z^2\right)+
\rho^2 \d\phi^2\right],
\end{equation}
where $U$, $A$, $k$ are functions of $\rho$ and $z$.
The axial Killing vector
is given by $\vec\eta=\partial_\phi$ and the axis of symmetry
is located at $\rho=0$.
The coordinate $t$ can be chosen to %have an intrinsic meaning, namely to
measure proper time of an observer at infinity, and hence
the Killing vector $\vec\xi= \partial_t$ is unit at infinity.
With this choice, the
remaining coordinate freedom in (\ref{eq:ds2e}) consists only of
constant shifts of $t$, $\phi$ and $z$.

Denoting by $F_{\alpha\beta}$ the e-m tensor
and by $\vkk$ a (non-null) Killing vector (will be either $\vec\xi$ or
$\vec\eta$), since $\lie(\vkk)F_{\alpha\beta}=0$
(stationary and axisymmetric e-m field)
there exist two scalar e-m potentials $E_{\kk},B_{\kk}$
with respect to $\vkk$ such that the
corresponding e-m fields $\vec E(\vkk)$ and $\vec B(\vkk)$
are given by
$E(\vkk)_\alpha(\equiv F_{\alpha\beta} \kk^\beta)=\partial_\alpha E_{\kk}$
and $B(\vkk)_\alpha(\equiv * F_{\alpha\beta} \kk^\beta)=
\partial_\alpha B_{\kk}$.
For convenience, one defines $\Lambda_{\kk}\equiv E_{\kk}+i B_{\kk}$,
which, given $\vkk$, contains all the information of the e-m field.
%FORMULA HERE FOR $F$?

It is well known (see e.g. \cite{sol})
that the Einstein-Maxwell equations
imply the existence of the so-called Ernst (complex) potential
with respect to $\vkk$
\begin{eqnarray}
\ernst_{\kk} = - N_{\kk} - \Lambda_{\kk} \bar\Lambda_{\kk} + i \Omega_{\kk},
\label{eq:ernst}
\end{eqnarray}
where $N_{\kk}\equiv \kk_\alpha \kk^\alpha$,
the bar indicates the complex conjugate,
and the real potential
$\Omega_{\kk}$ is such that
%another real potential $\Omega_{\kk}$
%(with respect to $\vkk$), such that
\begin{eqnarray}
  \d \Omega_{\kk}=\bm w_{\kk} - i \left(\Lambda_{\kk} \d \bar\Lambda_{\kk}-
\bar\Lambda_{\kk} \d \Lambda_{\kk}\right),
\label{eq:domega}
\end{eqnarray}
where $\bm w_{\kk}\equiv * (\bm \kk \wedge \d \bm \kk)$
is the twist 1-form of $\vkk$. % and the bar indicates the complex conjugate.
% Now, denoting by $N_{\kk}\equiv \kk_\alpha \kk^\alpha$ the norm of
% %the Killing vector
% $\vkk$, one introduces the so-called Ernst potential
% (with respect to $\vkk$) by
% \begin{eqnarray}
% \ernst_{\kk} = - N_{\kk} - \Lambda_{\kk} \bar\Lambda_{\kk} + i \Omega_{\kk}.
% \label{eq:ernst}
% \end{eqnarray}
The Einstein-Maxwell equations reduce then to an elliptic system
of equations for $(\ernst,\Lambda)$
% the potentials $\Lambda_k$ and $\ernst_k$,
%, for either $\kk=\vec \xi$ or $\kk=\vec \eta$,
known as the Ernst equations, plus a quadrature
in terms of $\ernst$ and $\Lambda$
for the function $k$. Dropping the $\kk$ indices,
the Ernst equations read
\begin{eqnarray}
  &&N \delta^{ij}\partial_i(\rho \partial_j \ernst)+
\rho\ \delta^{ij}\partial_i\ernst
\left(\partial_j\ernst + 2 \bar\Lambda\partial_j \Lambda\right)=0,
\label{eq:e1}\\
  &&N \delta^{ij}\partial_i(\rho \partial_j \Lambda)+
\rho\ \delta^{ij}\partial_i\Lambda
\left(\partial_j\ernst + 2 \bar\Lambda\partial_j \Lambda\right)=0,
\label{eq:e2}
\end{eqnarray}
where $N=-(\ernst+\bar\ernst+2\Lambda\bar\Lambda)/2$,
$i,j:\{\rho,z\}$,
and $\delta_{ij}$ is the $2\times 2$ identity. 
Given $(\ernst,\Lambda)$ solution of the Ernst equations,
all the information of the exterior electrovacuum solution
is recovered.
Choosing the form (\ref{eq:ds2e}),
apart from $k$, %for $\vkk=\vec\xi=\partial/\partial t$,
the metric for $\spE$ con be obtained by taking % and
$\vkk=\vec\xi=\partial_t$ so that
$U$ is obtained from $N_\xi=-e^{2U}$,
and $A$ is determined, up to a constant, by the quadrature
$\d A=-\rho N_\xi^{-2} *{}^{(h)}\bm w_\xi$, where
$*^{(h)}\d \rho=\d z, *^{(h)}\d z=-\d \rho$
and $\bm w_\xi$ has been obtained from (\ref{eq:domega}).
% \footnote{The additive constants in $A$ and $k$ will be determined
% from the values of the functions on $\sigma$ (see below and
% \cite{conv}).}

\section{The boundary conditions on $\sigma$}%the matching hypersurface}
The boundary for the exterior problem consists, in principle,
of the boundary associated to the surface of the body ($\sigma$,
whose spacelike cuts are topologically spheres) plus
the surface at infinity, where the flat asimptoticity
assumed on the exterior $\spE$ determines the behaviour
of the potentials $(\ernst,\Lambda)$ there.
This section is devoted to the boundary conditions
on the matching hypersurface
$\sigma$ that result from the matching conditions with a given
stationary and axisymmetric interior.

Given an interior, it has been proven~\cite{MASEuni,conv} that the matching
conditions fix $\sigma$ in the general case,
together with the values of the metric
functions $U$ and $A$ and their normal derivatives on $\sigma$:\footnote{
The additive constants in $A$ and $k$ are then determined, see \cite{conv}.
%${}^,$
%\footnote{
It has been assumed that the identification
of the exterior and interior across $\sigma$ has been prescribed,
in order to fix two extra degrees of freedom introduced
by the matching procedure, see \cite{MASEuni,mps}.}
\begin{equation}
  \label{eq:mc}
  U|_{\sigma},~~
  \vec n(U)|_{\sigma},~~
  A|_{\sigma},~~
  \vec n(A)|_{\sigma},
\end{equation}
where $\vec n$ denotes the normal to %the hypersurface
$\sigma$.
On the other hand, the continuity of $F_{\alpha\beta}$
%(because of the matching conditions for the e-m field)
\footnote{
It is assumed that no surface charges are present on the surface
of the body.} fix
\begin{equation}
  \label{eq:mcem}
  \d \Lambda|_{\sigma},
\end{equation}
for either $\vkk$,
so that in particular, $\Lambda|_{\sigma}$ is fixed up to
an additive arbitrary complex constant $\lambda$.

It is now straightforward to show that
the data given by (\ref{eq:mc}) translates
onto data for $\ernst$ %with respect to $\vkk=\vec\xi$
as follows. %(in the rest of the section $\vkk=\vec\xi$ is assumed
%and the corresponding $\kk$ indices are dropped).
% The form of %the matching hypersurface
% $\sigma$, with
% coordinates $\{\tau,\mu,\varphi\}$, can be always
% cast in the parametric form (see \cite{MASEuni,conv})
% $\sigma:\{t=\tau,\rho=\rho(\mu),z=z(\mu),\phi=\varphi\}$,
% and thus $\vec n=-\dot z\partial_\rho +\dot\rho\partial_z|_{\sigma}$,
% where the dot denotes differentitation with respect to $\mu$
% (so that $\dot f=\partial_\rho f \dot\rho+\partial_z f \dot z$).
% Therefore,
% %taking $\vkk=\vec\xi=\partial_t$ and dropping the $\kk$ indices,
% one has
% $
% %\begin{eqnarray*}
% \dot \Omega- i (\dot \Lambda \bar\Lambda-\dot{\bar\Lambda} \Lambda)|_{\sigma}=
% -\rho^{-1} N^2 \vec n(A)|_{\sigma}.
% %\end{eqnarray*}
% $
Using (\ref{eq:mc}), and taking into account $\lambda$ as given above,
the matching conditions fix $\Omega|_\sigma$ and $\d \Omega|_\sigma$
up to transformations of the form
 \begin{equation}
   \label{eq:mcomega}
   \Omega|_{\sigma}\rightarrow
   \Omega|_{\sigma}+c_\Omega -i(\bar\lambda \Lambda-
   \lambda\bar\Lambda)|_\sigma, \mbox{ and }\ 
   \d \Omega|_{\sigma}\rightarrow
   \d \Omega|_{\sigma}-i\left(\bar\lambda\d \Lambda-
     \lambda \d \bar\Lambda \right)|_\sigma,
 \end{equation}
where $c_\Omega$ is an arbitrary real constant.
% Analogously, the data (\ref{eq:mc}) fix
% $\vec n(\Omega)|_{\sigma}-i\left[\bar\lambda\vec n(\Lambda)-
% \lambda \vec n(\bar\Lambda)\right]$, which in turn, combined with
% (\ref{eq:mcomega}) implies that the matching conditions fix
% \begin{equation}
%   \label{eq:mcdomega}
%   \d \Omega|_{\sigma}-i\left(\bar\lambda\d \Lambda|_\sigma-
% \lambda \d \bar\Lambda|_\sigma \right).
% \end{equation}
From (\ref{eq:mc}),
%\footnote{In more generality,
%the $C^1$ continuity of the Killing vectors across $\sigma$
%ensures (\ref{eq:mcN}) for either $\vkk$ (see~\cite{mps})},
given an interior, the matching conditions fix %(for $\vkk=\vec\xi$)
\begin{equation}
  \label{eq:mcN}
  N|_{\sigma},~~
  \vec n(N)|_{\sigma}.
\end{equation}
The data for $\ernst|_{\sigma}$ and $\d \ernst|_{\sigma}$
can be obtained now from (\ref{eq:mcomega}) and (\ref{eq:mcN}).

Notice
that due to the elliptic character of the Ernst equations system for
$(\ernst,\Lambda)$,
the data coming from the matching conditions (Cauchy data)
overdetermines the problem.
Nevertheless, there is still three degrees of freedom
driven by $\lambda$ and $c_\Omega$ on that data.
The questions to address at this point are then:
(a) uniqueness of the exterior solution $(\ernst,\Lambda)$
given Dirichlet data on $\sigma$, i.e. $\ernst|_\sigma,\Lambda|_\sigma$,
and (b) if a solution $(\ernst,\Lambda)$ exsists, are
 $\lambda$ and $c_\Omega$ determined, and thus the Dirichlet data?
Of course, the affirmative answer to both questions
solves the issue of the uniqueness of the exterior field.

\section{Uniqueness given Dirichlet data on $\sigma$}
The proof, presented here schematically,
follows closely those used in the uniqueness theorems
of black holes (see \cite{heusler}), making use of
the very rich intrinsic structure of the Ernst equations.
In what follows, the $E$ suffix (for 'exterior') and
the $\kk$ indices will be omitted for simplicity.
Equations (\ref{eq:e1})-(\ref{eq:e2}) can be interpreted as
the Euler-Lagrange equations for the action \cite{heusler}
\[
{\cal S}=4\int\left(\frac{1}{4N^2}
|\d\ernst+2\bar\Lambda\d\Lambda|^2+\frac{1}{N}|\d\Lambda|^2\right)
\rho\d\rho\d z,
\]
where $|\bm \theta|^2\equiv g^{\alpha\beta}\theta_\alpha\bar\theta_\beta$.
%($E$ suffix has been dropped and indices $\kk$ are omitted for simplicity).
Taking $\Phi$ to be a hermitian $su(2,1)$ matrix defined
by ($a,b:1,2,3$)
\[
\Phi_{ab}\equiv \eta_{ab}+2\ \sign(N)\bar v_a v_b\]
where $\eta_{ab}=\mbox{diag}(-1,1,1)$ and
$
v_a=(2\sqrt{|N|})^{-1}(\ernst-1,\ernst+1,2\Lambda),
$
%so that $\eta^{ab}\bar v_a v_b=-\sign(N)$,
one can define a $su(2,1)$-valued 1-from by
\[
\bm\jj\equiv \Phi^{-1}\cdot \d \Phi.
\]
The pairs of solutions $(\ernst,\Lambda)$ have been now translated onto
$\Phi$.
In terms of $\jj$, the above action is rewritten as \cite{heusler}
\[
{\cal S}=\int\frac{1}{2}
g_{\alpha\beta}\Tr\left(\jj^\alpha\cdot\jj^\beta\right)
\rho\d\rho\d z,
\]
%($\Tr$ stands for the trace)
for which the variational equation reads \cite{heusler}
\begin{equation}
  \label{eq:jjdiv}
  \nabla_\alpha\jj^\alpha=0.
\end{equation}
The key property for the proofs of the uniqueness theorems
is the positivity of $\Phi$ (and the action), which is ensured by taking
$\vkk=\vec\eta$.
Let us define $X\equiv N_\eta=e^{-2U}\rho^2-e^{2U}A^2(>0)$,
$Y\equiv \Omega_\eta$
and $A_\eta\equiv -Ae^{2U} X^{-1}$. The line-element
(\ref{eq:ds2e}) can be cast as
\[
\d s^2=-X^{-1}\rho^2 \d t^2+ X(\d \phi+A_\eta\d t)^2+X^{-1}e^{2k}
(\d \rho^2+\d z^2).
\]
The problem in this choice is that $N$, i.e. $X$, vanishes
on the axis, and therefore a careful analysis there is needed
(see \cite{carterbh}). It is convenient to change to
prolate spheroidal coordinates $\{x,y\}$, $x>1,|y|\leq 1$
by $\rho^2=\nu^2(x^2-1)(1-y^2),
z=\nu x y$ ($\nu>0$) so that $\d\rho^2+\d z^2=\nu^2(x^2-y^2)
\widetilde{\d s^2}$
where $\widetilde{\d s^2}=\frac{1}{x^2-1}\d x^2+\frac{1}{1-y^2}\d y^2\equiv
h_{ij}\d x^i\d x^j$. The boundaries to the domain $D$ for the solutions
%we are interested in
are now given as follows: (i) infinity, located at $x=\infty$,
(ii) the two parts of the axis at $y=\pm 1$,
(iii) the boundary corresponding to
$\sigma$, a curve $\Sigma$ with $x>1$ joining $y=1$ with $y=-1$.

The core of the proof consists in
considering two sets of solutions $\Phi_{(1)}$ and
$\Phi_{(2)}$, with corresponding pairs $\bm\jj_{(1)},\bm\jj_{(2)}$,
$X_{(1)},X_{(2)}$, etc, with common Dirichlet %boundary
data on~$\Sigma$.
Defining $\underline f\equiv f_{(1)}-f_{(2)}$
and $\Psi\equiv \Phi_{(1)}\cdot \Phi_{(2)}^{-1}-\id$,
where $\id$ is the identity,
%so that $\Psi=0\Leftrightarrow \Phi_{(1)}=\Phi_{(2)}$.
and taking into account that $\bm\jj$ has only components in $\rho,z$
($i,j$),
the Mazur identity follows (see \cite{carterbh,heusler}):
\[
(\rho \Tr \Psi^{;i})_{;i}=\rho h_{ij}\Tr\{\jjj^{\dagger i}\cdot\Psi_{(2)}
\cdot \jjj^j\cdot \Psi_{(1)}\}.
\]
If $\bm\jjj\neq 0$, and since $\rho$, $h_{ij}$ and $\Phi$
are positive definite, then $(\rho \Tr \Psi^{;i})_{;i}>0$.

%By using the Stokes theorem,h
The point here lies on showing that the data
on the boundary of our domain $D$, $\partial D$, implies
that the integral over $\partial D=(i)+(ii)+\Sigma$ of
$\rho \Tr\Psi^{;i}\d S_i$ vanish, and use then the Stokes theorem.
Explicitly, defining $\widetilde f\equiv f_{(1)}+f_{(2)}$,
one has \cite{heusler}
\begin{equation}
  \label{eq:trace}
  \Tr\Psi=\frac{1}{X_{(1)}X_{(2)}}
  \left[\underline X^2+2 \widetilde X |\underline\Lambda|^2+
    |\underline\Lambda|^4
  +\left[\underline Y+
    \Ima(\underline \Lambda\bar{\widetilde\Lambda})\right]^2\right].
\end{equation}
Following the black hole theorems,
taking the values of $X$, $Y$ and $\Lambda$ on the axis
and the limits at infinity
as were computed by Carter (see \cite{carterbh,heusler}),
expression (\ref{eq:trace}) leads
to the vanishing of $\rho \Tr \Phi^{;i}$ on (i) and (ii).
Now, since the numerator
of $\Tr\Psi$ is quadratic in $\underline X$, etc,
and by assumption $\underline \Phi|_{\Sigma}=0$, one infers
$\Tr \Psi_{;i}|_{\Sigma}=0$.

Therefore $\jjj=0$. Since by construction
$\rho\Psi^{;\alpha}=\Psi_{(2)}\cdot \jjj^\alpha\cdot \Phi_{(1)}^{-1}$,
$\Psi$ is constant all over $D$, and thus $\Psi=0$ because $\Psi|_{\Sigma}=0$
by assumption. This ends the proof showing that given Dirichlet data
on $\sigma$, $(\ernst_\eta|_{\sigma},\Lambda_\eta|_\sigma)$,
the solution $(\ernst_\eta,\Lambda_\eta)$ of the Ernst
equations in the exterior region $\spE$ is unique.
Finally, the correspondence between conjugate solutions
$(\ernst_\eta,\Lambda_\eta)$ and $(\ernst_\xi,\Lambda_\xi)$
(see~\cite{heusler}) leads to the same result for $\vkk=\vec\xi$. 

\section{Fixing the Dirichlet data}
%The Dirichlet data on $\sigma$ has three degrees of freedom, driven
%by $c_\Omega$ and $\lambda$.
The purpose now
is to show how the full set of Cauchy
data, i.e. taking into account
$(\vec n(\ernst)|_\sigma,\vec n(\Lambda)|_\sigma)$,
fixes the values of $c_\Omega,\lambda$,
provided that the solution exists. % (the problem is overdetermined).

Following \cite{MASEuni}, the proof makes use of the
divergence free fields $\bm\jj$ (\ref{eq:jjdiv}) choosing $\vkk=\vec\xi$.
Decomposing $\bm\jj$ on a basis of $su(2,1)$, 8 conserved
1-forms are obtained, which are used to define
a divergence free real 1-form in $\spE$ depending on 8 real constants
$
%\begin{eqnarray*}
%&&
\bm W(\Lambda,\ernst;c_e,c_k,c_d,c_{h_1},c_{h_2},c_{a_1},c_{a_2},c_s)
\equiv
%\\
%&&~~~~
c_e\bm\jj^e+c_k\bm\jj^k+c_d\bm\jj^d+
c_{h_1}\bm\jj^h+c_{h_2}\bar{\bm\jj}^h+
c_{a_1}\bm\jj^a+c_{a_2}\bar{\bm\jj}^a+
c_s\bm\jj^s,
%\end{eqnarray*}
$
where all $\bm\jj=\bm\jj(\ernst,\Lambda)$ and such that
the only non-vanishing surface integrals at infinity are
$\int_{S_\infty}\jj_\alpha^d \d S^\alpha=8\pi M$,
$\int_{S_\infty}\jj_\alpha^h \d S^\alpha=-4\pi q$
and
$\int_{S_\infty}\jj_\alpha^a \d S^\alpha=-4\pi q$,
where $M(\ernst,\Lambda)$ and the complex $q(\ernst,\Lambda)$
relative to a given solution $(\ernst,\Lambda)$
correspond to the mass and the e-m charge of the configuration,
respectively.
As a consequence of $W^\alpha{}_{;\alpha}=0$, we have
\begin{equation}
\int_{\Sigma}W_\alpha \d S^\alpha=
\int_{S_\infty}W_\alpha \d S^\alpha=8\pi M c_d
-4\pi \left[q(c_{h_1}+c_{a_1})+\bar q (c_{h_2}+c_{a_2})\right].
\label{eq:int}
\end{equation}
Now, we assume two exterior solutions for the same
interior exist $(\ernst_{(1)},\Lambda_{(1)})$,
$(\ernst_{(2)},\Lambda_{(2)})$ such that their Cauchy
boundary data differs by $\lambda$ and $c_\Omega$,
i.e.
$\Lambda_{(1)}|_\sigma=\Lambda_{(2)}|_\sigma+\lambda$,
$\d \Lambda_{(1)}|_\sigma=\d \Lambda_{(2)}|_\sigma$,
$N_{(1)}|_\sigma=N_{(2)}|_\sigma$,
$\d N_{(1)}|_\sigma=\d N_{(2)}|_\sigma$,
$\Omega_{(1)}|_{\sigma}=
\Omega_{(2)}|_{\sigma}-c_\Omega +i(\bar\lambda \Lambda_{(2)}-
   \lambda\bar\Lambda_{(2)})|_\sigma$,
$\d \Omega_{(1)}=\d \Omega_{(2)}|_{\sigma}
+i\left(\bar\lambda\d \Lambda_{(2)}-
     \lambda \d \bar\Lambda_{(2)} \right)|_\sigma$.
From this, the following equality holds on $\sigma$
\begin{equation}
\bm W(\ernst_{(1)},\Lambda_{(1)};c_e,\ldots,c_s)|_\sigma=
\bm W(\ernst_{(2)},\Lambda_{(2)};\hat c_e,\ldots,\hat c_s)|_\sigma,
\label{eq:ws}
\end{equation}
for a set of 8 certain explicit relations
\begin{eqnarray*}
  &&\hat c_e=\hat c_e(c_e,\ldots, c_s; c_\Omega,\lambda),~~~
%   \vspace{-5mm}\\
%   &&\vspace{-5mm}\vdots\\
%  &&
\ldots,~~~\hat c_s=\hat c_s(c_e,\ldots, c_s; c_\Omega,\lambda).
\end{eqnarray*}
The integration of (\ref{eq:ws}) over $\sigma$ using (\ref{eq:int})
leads to
\begin{eqnarray*}
  &&8\pi M_{(1)} c_d
-4\pi \left[q_{(1)}(c_{h_1}+c_{a_1})+\bar q_{(1)} (c_{h_2}+c_{a_2})\right]
\\
&&~~~~ = 8\pi M_{(2)} \hat c_d
-4\pi \left[q_{(2)}(\hat c_{h_1}+\hat c_{a_1})+
\bar q_{(2)} (\hat c_{h_2}+\hat c_{a_2})\right],
\end{eqnarray*}
which has to hold for arbitrary choices of the 8 constants
$\{c_e,\ldots,c_s\}$.
A straightforward calculation leads to the fact that
\[
M_{(1)}+M_{(2)}\neq 0 \Rightarrow \lambda=c_\Omega=0.
\]
For physically well behaved solutions the total
mass should be positive, and thus this result implies that
both $\lambda$ and $c_\Omega$ must vanish,
and hence the uniqueness of the exterior solution
generated by a given interior distribution of matter
in stationary and axially symmetric rotation follows.

I wish to thank Brien Nolan for reading this manuscript.
This work was produced while I was in Queen Mary, University
of London, and funded by the EPSRC grant GR/R53685/01.
I also thank the IRCSET for grant PD/2002/108.

%in these coordinates, Carter (see \cite{carterbh}) computed
%the asymptotic behaviour of $X$, $Y$ and $\Lambda$ as $x\rightarrow\infty$

%\section{References}

\end{document}